\documentclass[preprint,12pt]{aastex}
\usepackage{natbib}
\newcounter{address}
\newcommand{\latin}[1]{\textit{#1}}
\newcommand{\ie}{\latin{i.e.}}

\begin{document}

\title{A more informative picture of the HST Ultra Deep Field}
\author{
  Nicholas~Wherry\altaffilmark{\ref{NYU}},
  Michael~R.~Blanton\altaffilmark{\ref{NYU}}
  David~W.~Hogg\altaffilmark{\ref{NYU},\ref{email}}
}
\setcounter{address}{1}
\altaffiltext{\theaddress}{\stepcounter{address}\label{NYU}
Center for Cosmology and Particle Physics, Department of Physics, New
York University, 4 Washington Pl, New York, NY 10003}
\altaffiltext{\theaddress}{\stepcounter{address}\label{email}
\texttt{david.hogg@nyu.edu}}

\begin{abstract}
The Hubble Space Telescope has obtained some of its well-deserved
impact by producing stunning three-color (RGB) pictures from
three-band imaging data.  Here we produce a new RGB representation of
the $I$, $V$, and $B$ bandpass images of the HST Ultra Deep Field
(HUDF).  Our representation is based on principles set forth elsewhere
(Lupton et al 2004, PASP, 116, 133--137).  The principal difference
between our RGB representation of the data and the more traditional
representation provided by the Space Telescope Science Institute is
that the (necessarily) nonlinear transformation between data values
and RGB values is done in a color-preserving (\ie, hue- and
saturation-preserving) way.  For example, if one of the image pixel
values saturates the dynamic range of the RGB representation, all
three of the R, G, and B values in the representation are truncated
such that the hue and saturation of the pixel is the same as it would
have been if the pixel had had lower flux but the same astronomical
color.  This, in effect, makes the bright parts of the representation
an informative color map, not ``whited out'' as they are in
traditional representations.  For the HUDF, this difference is seen
best in the centers of bright galaxies, which show significant detail
not visible in the more traditional representation.
\end{abstract}

\begin{figure}
\plotone{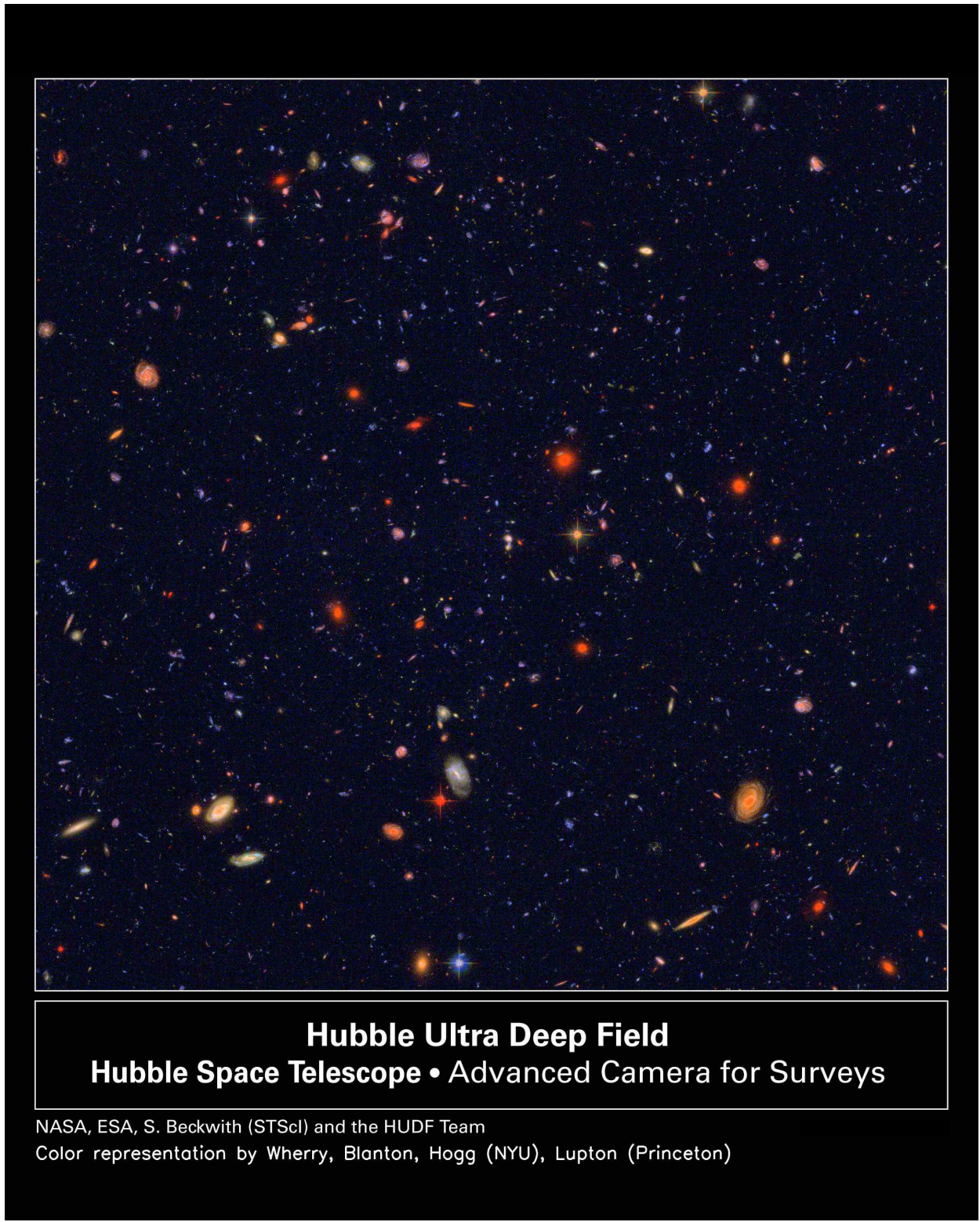}
\caption[]{The new representation.  Also available as full resolution
JPEG image from ``http://cosmo.nyu.edu/hogg/visualization/''.}
\end{figure}

\begin{figure}
\plotone{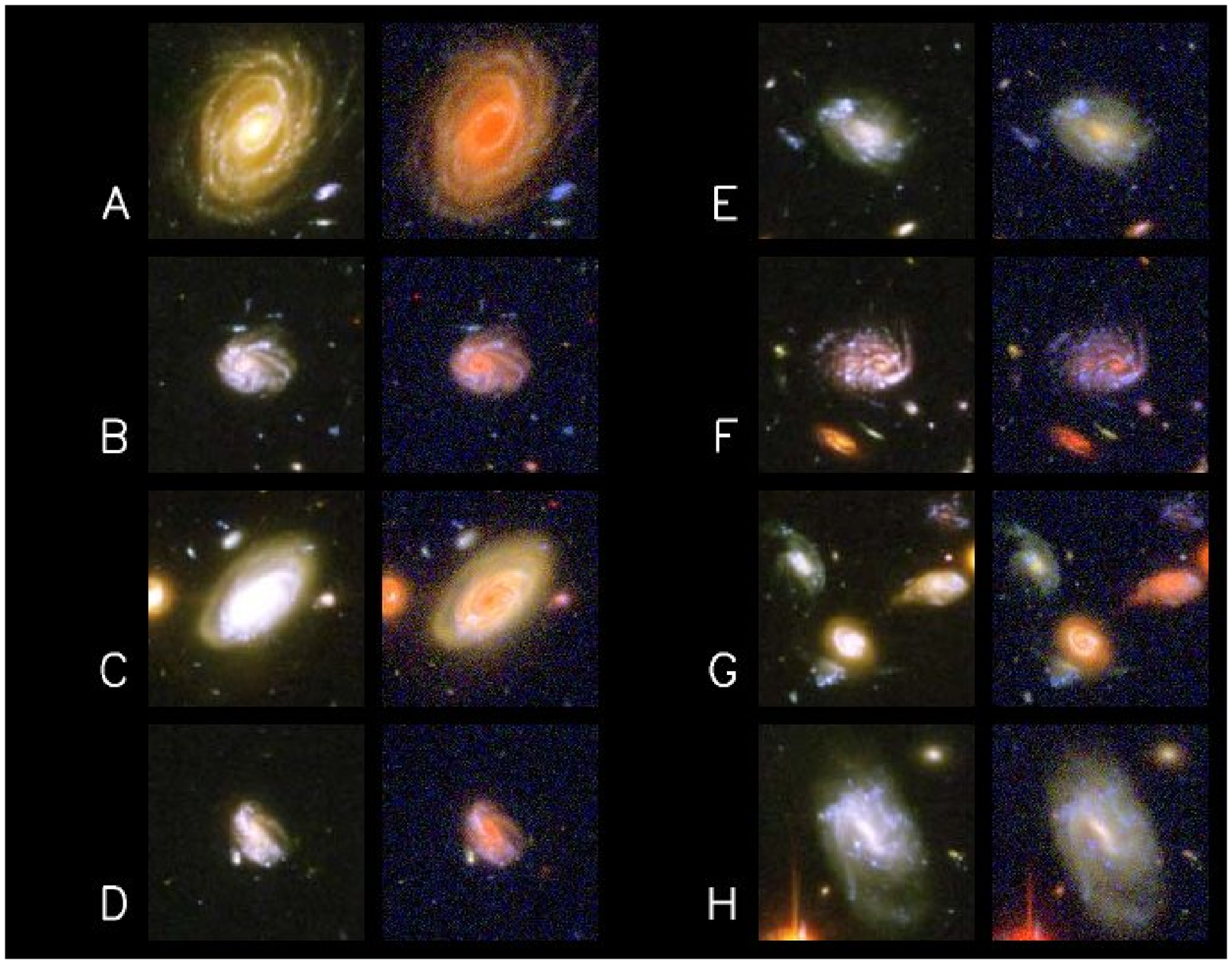}
\caption[]{The traditional and new color representations compared,
with traditional on the left and new on the right.  The background
looks bluer in the new representation because we have not done
anything to suppress the higher noise in the blue ($B$ band) image.
Brighter galaxies look redder in the new representation primarily
because the centers of the galaxies are saturated to their true color
rather than white.  \textsf{A}, \textsf{B}, \textsf{C}, and \textsf{D}
have red centers which are obvious in the new image.  \textsf{C},
\textsf{E}, and \textsf{H} are lower redshift galaxies, which appear
to be composed of ``pastel'' colors while \textsf{B}, \textsf{D} and
\textsf{F} are higher redshift galaxies, which display brighter, more
saturated colors because the 4000~\AA\ break appears in the visual
band.  The triple in \textsf{G} shows much more color diversity in the
new representation.  \textsf{E} and \textsf{H} have low surface
brightness outer features and high surface brightness central features
that are simultaneously visible.}
\end{figure}

\acknowledgements It is a pleasure to thank Steve Beckwith and the
HUDF team for taking, reducing, and making public the HUDF data.  We
have benefitted from advice and software from Doug Finkbeiner, Robert
Lupton, and David Schlegel.  Financial support for this project was
provided by NASA grant NAG5-11669, an REU extension to NSF grant
PHY-0101738, and the NYU Dean's Undergraduate Research Fund.

\end{document}